\documentclass[preprint,12pt]{article}
\usepackage{epsfig}
\usepackage{lineno}
\usepackage{rotating}
\usepackage{booktabs}
\usepackage{lscape}
\usepackage{amsmath}
\usepackage{multirow}
\usepackage[usenames]{color}
\usepackage{amssymb}

\usepackage{cite}

\newcommand\TopSpace{\rule{0pt}{2.6ex}}
\newcommand\BottomSpace{\rule[-0.9ex]{0pt}{0pt}}

\usepackage[footnotesize]{caption} 
\begin{document}

    \begin{center}
        \vspace*{1.0cm}

        {\Large \bf{Search for $\alpha$ decay of naturally occurring Hf-nuclides using a Cs$_2$HfCl$_6$ scintillator}}
        \vspace*{1.0cm}

        {\bf
        V.~Caracciolo$^{a,b,c}$\footnote{Corresponding author. {\it E-mail address:} vincenzo.caracciolo@roma2.infn.it.},
        S.S.~Nagorny$^d$,
        P.~Belli$^{a,b}$,
        R.~Bernabei$^{a,b,}$,
        F.~Cappella$^{e,f}$,
        R.~Cerulli$^{a,b}$,
        A.~Incicchitti$^{e,f}$,
        M.~Laubenstein$^c$,
        V.~Merlo$^{a,b}$,
        S.~Nisi$^c$,
        P.~Wang$^g$
        }

    \end{center}

        \vskip 0.3cm

        {\footnotesize\noindent
        $^a$ INFN sezione di Roma ``Tor Vergata'', I-00133 Rome, Italy.\\
        $^b$ Dipartimento di Fisica, Universit\`a di Roma ``Tor Vergata'', I-00133, Rome, Italy.\\
        $^c$ INFN, Laboratori Nazionali del Gran Sasso, I-67100 Assergi (AQ), Italy.\\
        $^d$ Department of Physics, Queen's University, Kingston, ON K7L 3N6, Canada.\\
        $^e$ INFN sezione di Roma, I-00185 Rome, Italy.\\
        $^f$ Dipartimento di Fisica, Universit\`a di Roma ``La Sapienza'', I-00185 Rome, Italy.\\
        $^g$ Department of Chemistry, Queen's University, Kingston, ON K7L 3N6, Canada.\\
        }

\begin{abstract}
    Residual radioactive contaminants of a caesium hafnium chloride (Cs$_2$HfCl$_6$) crystal scintillator have
    been measured in a low background setup at the Gran Sasso National Laboratory of the INFN, Italy.
    The total alpha activity  of the detector is at the level of 7.8(3)~mBq/kg.
    The results of direct studies  of the $\alpha$ decay of naturally occurring  Hf~isotopes that have been performed using  the ``source=detector"
    approach  are presented.
    In 2848~h of data taking, the $\alpha$ decay of $^{174}$Hf was observed with T$_{1/2}~=~(7.0\pm1.2)\times10^{16} $~y.
\end{abstract}

\vskip 0.4cm


\section{Introduction}
Since its discovery more than a century ago,  $\alpha$ decay remains one of the most powerful tools to study
 nuclei and their structure.  $\alpha$ decay is energetically favourable for naturally occurring heavy isotopes (from $^{142}$Ce to
$^{238}$U), while the probability of the $\alpha$ particles tunnelling through the nuclear potential barrier is
significantly reduced for nuclei lighter than the $^{209}$Bi.
In fact, these light nuclei --
 their available decay energy is less than 3 MeV -- undergo $\alpha$ emission
with  very long half-life ($\gtrsim 10^{14}$ y); therefore,
these processes are extremely difficult to  detect with conventional techniques.
However, significant progress has been made in the field of search for  rare $\alpha$ decays during the past
decade. This was mostly triggered by the development and application of new experimental techniques, as well as by
 improving  well-known ones. A recent detailed review on investigations
 of  rare $\alpha$ decays and  the achieved results is given in Ref. \cite{epjareview}.
It should be noted that for some $\alpha$ active elements there are still no detector
materials that may contain  the element of interest in an amount significant enough to perform highly sensitive
measurements with the ``source = detector" approach.

Very recently  there has been a  significant renewed interest in crystal scintillators  as the K$_2$PtCl$_6$ \cite{Meija}; in fact, their properties include: a high light yield, a very good linear response at low
energies and a good energy resolution.
The Cs$_2$HfCl$_6$ (CHC) crystal -- belonging to the same structure group -- is one of the promising new
scintillating materials for $\gamma$ spectroscopy offering a light output of more than 50000 photons/MeV,
a 3.3\% energy resolution
at 662 keV \cite{Beeman13}, with an excellent ability for pulse shape discrimination (PSD) of the $\gamma$(e)/$\alpha$ scintillation  signals
\cite{nagornyNIMA}. Moreover, this is also the first
scintillating material containing a high fraction of Hf ($\sim$ 27\% in mass)  that can be easily produced using
the Bridgman growing technique.

First investigations of the chemical purity of a CHC crystal using inductively coupled plasma mass spectrometry (ICP-MS)
 and of its radio-purity using an ultra-low background high purity germanium (HP-Ge) detector have shown that this compound is
clean with respect to  the U/Th radioactive chains (only limits were set at the level of few mBq/kg). However,
some contamination with man-made $^{137}$Cs ($\sim$ 1 Bq/kg) and $^{134}$Cs ($\sim$ 50 mBq/kg), and cosmogenic
$^{132}$Cs ($\sim$ 25 mBq/kg) and $^{181}$Hf ($\sim$ 15 mBq/kg)
was measured in the sample \cite{nagornyNIMAcontaminanti}.
Thus, due to its promising good radio-purity and to
the combination of its outstanding scintillation features, the CHC crystal opens new possibilities in the
search for the rare nuclear processes  in natural Hf.

In this paper, we show the successful operation of a CHC crystal as  low-background scintillator
to investigate  rare $\alpha$ decays of naturally occurring Hf isotopes,  specifically of  $^{174}$Hf.

We remind that the first and only measurement of the $^{174}$Hf $\alpha$ decay was
carried out more than fifty years ago  using an ionization chamber with thin and low mass  HfO$_2$ samples. Only a preliminary
indication of its detection was obtained due to an extremely low signal-to-background ratio, quoting as half-life the value
 2.0(4) $\times$ 10$^{15}$ y \cite{Macfarlane61} while theoretical predictions in various models quote a
 half-life value ranging  from 3.5 $\times$ 10$^{16}$ y to 7.4 $\times$ 10$^{16}$ y \cite{Abuck91,poe83,Teo3}, i.e. one order of magnitude higher.

\begin{figure}[!ht]
  \centering
  \includegraphics[width=0.8\textwidth]{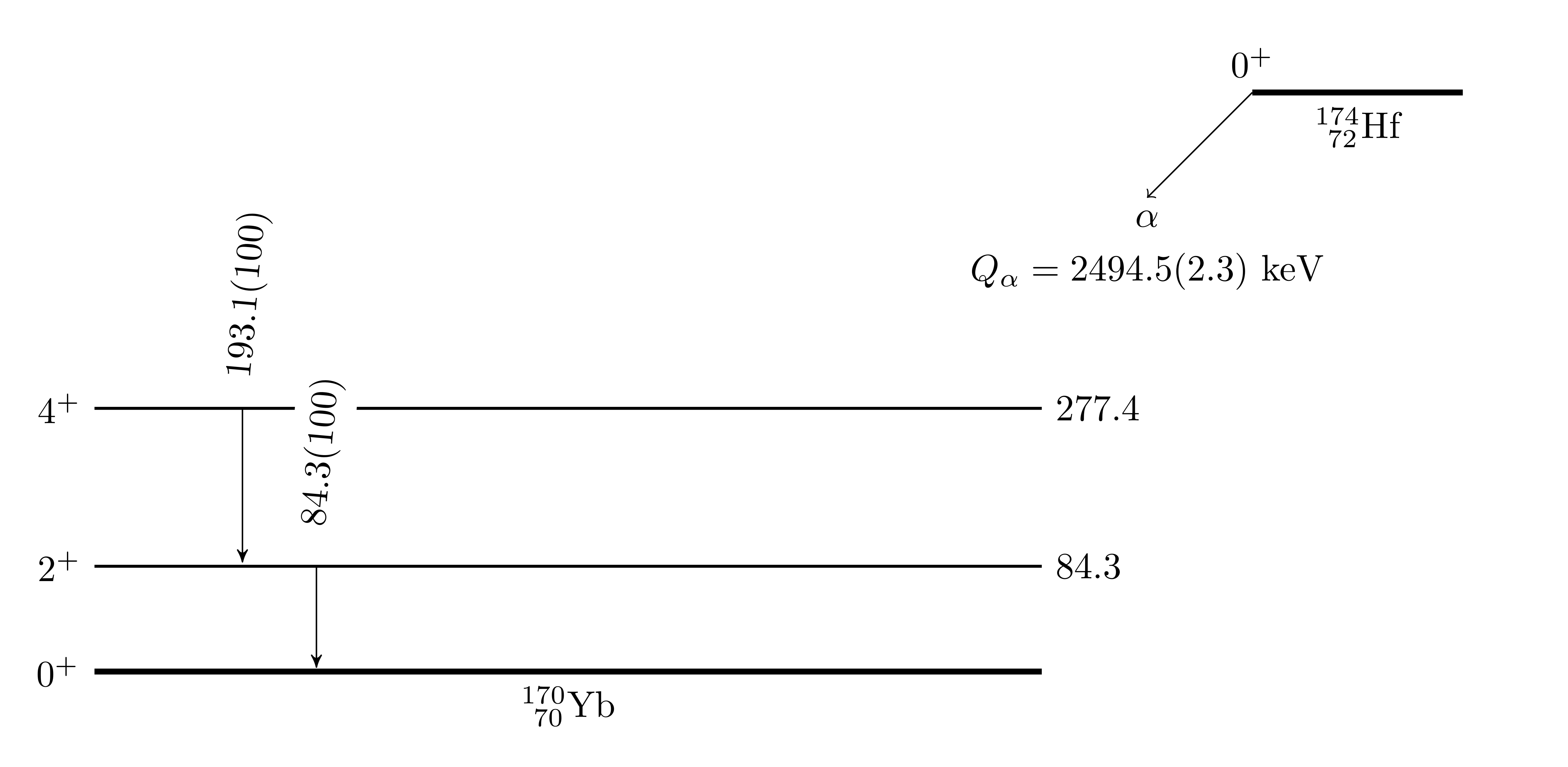}
  \caption{Simplified scheme of the $\alpha$ decay of  $^{174}_{~72}$Hf to $^{170}_{~70}$Yb. The energies of the excited levels and of the emitted $\gamma$ quanta are in keV (the relative intensity of the $\gamma$ quantum from each level
is given in parentheses). The $\alpha$ decay schemes of all the naturally occurring Hf isotopes are reported in
 Ref. \cite{danevichchc19}.}\label{fig:schemalivello}
\end{figure}
\begin{sidewaystable}
\begin{center}
\caption{Some potential $\alpha$ transitions of Hf isotopes and related information. Only naturally occurring isotopes (with natural abundance $\delta$) and with Q$_\alpha>0$ between  g.s. transitions or between g.s. and lowest bound level transitions (with spin/parity $J^{\pi}$) are listed. E$_\alpha$ is the kinetic energy of the
alpha particle. N is the number of nuclei in the CHC crystal used in this work.
Experimental measurements (when available) and theoretical prediction of the half-live are reported in the
last four columns. T$_{1/2}$ values have been calculated in Refs.~\cite{epjareview,danevichchc19}, following  the prescription in Refs.~\cite{Abuck91,poe83,Teo3}.}
\resizebox{\linewidth}{!}{%
\begin{tabular}{ c | c | c | c  | c | c | c | c c c }
\hline
\hline
\multirow{1}*{~}                  &        ~                &    ~       &    ~          &    ~      &      ~          &  \multicolumn{4}{c }{~}                                                         \\
                Nuclide           &   $J^{\pi}$             & $\delta$   & Q$_\alpha$    & E$_\alpha$&      N          &  \multicolumn{4}{c }{T$_{1/2}$ (y)}                                            \\
                Transition        &       of                &    ~       &    ~          &    ~      &      ~          &  \multicolumn{4}{c }{~}                                                         \\
 \cline{7-10}       ~             &  Parent $\rightarrow$   &   (\%)     & (keV)         & (keV)     &      ~          &           ~              &      ~            &Theoretical\cite{epjareview,danevichchc19}      &    ~              \\
                                  &  Daughter Nuclei        &\cite{Meija}& \cite{2017Wa10} &   ~       &      ~          &       Experimental       & \cite{Abuck91}       &  \cite{poe83}    &  \cite{Teo3}     \\
                    ~             &  and its level (keV)\cite{2012Wa38,2017Wa10}    &    ~       &    ~          &   ~       &      ~          &         ~                &       ~           &     ~           &     ~            \\
 \hline\TopSpace\BottomSpace
\multirow{2}*{$^{174}$Hf$~\rightarrow~^{170}$Yb} & $0^+\rightarrow0^+$, g.s. &\multirow{2}*{0.16(12)} & \multirow{2}*{2494.5(2.3)}& \multirow{2}*{2437.6(2.2)}&\multirow{2}*{$1.0\times10^{19}$}&$2.0(4)\times10^{15}$ \cite{Macfarlane61,T174hf02} & $3.5\cdot10^{16}$ & $7.4\times10^{16}$& $3.5\times10^{16}$\\
& $0^+\rightarrow2^+$, 84.2 & & &           && $\geqslant 3.3\cdot10^{15}$\cite{danevichchc19} & $1.3\cdot10^{18}$  & $3.0\times10^{18}$& $6.6\times10^{17}$\\
\hline\TopSpace\BottomSpace
\multirow{2}*{$^{176}$Hf$~\rightarrow~^{172}$Yb} & $0^+\rightarrow0^+$, g.s. &\multirow{2}*{5.26(70)}   & \multirow{2}*{2254.2(1.5)}& \multirow{2}*{2203.3(1.5)}       &\multirow{2}*{$3.3\times10^{20}$}& --                     &  $2.5\times10^{20}$& $6.6\times10^{20}$& $2.0\times10^{20}$\\
 & $0^+\rightarrow2^+$, 78.7 &   & &        &&                     $\geqslant3.0\times10^{17}$ \cite{danevichchc19} &  $1.3\times10^{22}$& $3.5\times10^{22}$& $4.9\times10^{21}$\\
\hline\TopSpace\BottomSpace
\multirow{2}*{$^{177}$Hf$~\rightarrow~^{173}$Yb} & $7/2^-\rightarrow5/2^-$, g.s. &\multirow{2}*{18.60(16)}   & \multirow{2}*{2245.7(1.4)}& \multirow{2}*{2195.3(1.4) }      &\multirow{2}*{$1.2\times10^{21}$}& --                       &  $4.5\times10^{20}$& $5.2\times10^{22}$& $4.4\times10^{22}$\\
& $7/2^-\rightarrow7/2^-$, 78.6 & & &&& $\geqslant 1.3\times10^{18}$   \cite{danevichchc19}                    &  $9.1\times10^{21}$& $1.2\times10^{24}$& $3.6\times10^{23}$\\
\hline\TopSpace\BottomSpace
\multirow{2}*{$^{178}$Hf$~\rightarrow~^{174}$Yb} & $0^+\rightarrow0^+$, g.s.    &\multirow{2}*{27.28(28)}   & \multirow{2}*{2084.4(1.4)}& \multirow{2}*{2037.9(1.4) }      &\multirow{2}*{$1.7\times10^{21}$}& --                       &  $3.4\times10^{23}$& $1.1\times10^{24}$& $2.2\times10^{23}$\\
& $0^+\rightarrow2^+$, 76.5     & & & && $\geqslant 2.0\times10^{17}$   \cite{danevichchc19}                    &  $2.4\times10^{25}$& $8.1\times10^{25}$& $7.1\times10^{24}$\\
\hline\TopSpace\BottomSpace
\multirow{2}*{$^{179}$Hf$~\rightarrow~^{175}$Yb} & $9/2^+\rightarrow7/2^+$, g.s. &\multirow{2}*{13.62(11)}   & \multirow{2}*{1807.7(1.4)}& \multirow{2}*{1767.6(1.4)} &\multirow{2}*{$8.6\times10^{20}$}& $\geqslant 2.2\times10^{18}$ \cite{danevichchc19}                       &  $4.5\times10^{29}$& $4.0\times10^{32}$& $4.7\times10^{31}$\\
& $9/2^+\rightarrow9/2^+$, 104.5&   & & && $\geqslant2.2 \times10^{18}$ \cite{danevichchc19}                       &  $2.0\times10^{32}$& $2.5\times10^{35}$& $2.2\times10^{34}$\\
\hline\TopSpace\BottomSpace
\multirow{2}*{$^{180}$Hf$~\rightarrow~^{176}$Yb} & $0^+\rightarrow0^+$, g.s.     &\multirow{2}*{35.08(33)}   & \multirow{2}*{1287.1(1.4)}& \multirow{2}*{1258.7(1.4)}&\multirow{2}*{$2.2\times10^{21}$}& --                       &  $6.4\times10^{45}$& $5.7\times10^{46}$& $9.2\times10^{44}$\\
& $0^+\rightarrow2^+$, 82.1     &   & &  && $\geqslant 1.0\times10^{18}$  \cite{danevichchc19}                     &  $4.0\times10^{49}$& $4.1\times10^{50}$& $2.1\times10^{48}$\\
\hline
\hline
\end{tabular}
\label{t:Q}}
\end{center}
\end{sidewaystable}
In order to detect such  rare decays, it is necessary to use radio-pure  materials  to
maximize the signal-to-background
ratio. Thus, we give in this work,   the chemical purity and  the radio-active
contamination of the used  CHC crystal ``as is" in the light of future applications in low-background experiments to investigate
rare processes in Hf.
Information about some $\alpha$ transitions of naturally occurring  Hf isotopes  with Q$_\alpha>0$  are listed in Table  \ref{t:Q}.
All the naturally occurring Hf isotopes can decay to  g.s. or to excited levels of their daughter.
Fig. \ref{fig:schemalivello} shows a simplified expected decay scheme of the $\alpha$ decay of the $^{174}$Hf isotope;
the expected $\alpha$ decay schemes of all
the naturally occurring Hf isotopes are reported in Ref. \cite{danevichchc19}.

\section{The experiment}\label{p:exp}

\subsection{The CHC crystal scintillator}
The growth of the CHC crystal was done using the Bridgman technique. The
initial compounds were from a commercially available feedstock with the highest available
chemical purity of CsCl beads and HfCl$_4$ powder. The purity level of CsCl beads was 99.998\%.
To reach a similar purity grade of HfCl$_4$  powder, the compound was  processed by
three-fold purification  (see e.g. Ref. \cite{nagornyNIMA}). Later, to prepare the CHC compound for the
crystal growth, the stoichiometric mixture of purified HfCl$_4$ powder and
CsCl beads was well mixed  and placed into a quartz ampule sealed under
vacuum. Then, the ampule was transferred to a two-zone vertical furnace to perform the
crystal growth with a crystallization rate of 1 cm/day and applying a thermal
gradient of 5$^{\circ}$C/cm at the  solid/liquid  interface. A more detailed
description of the purification of the starting material  and the CHC crystal growth is given  in Ref. \cite{nagornyNIMA}. A CHC crystal sample with a mass of 6.90(1)~g,
 22 mm diameter and 4.6 mm height ($\rho\sim3.9$~g/cm$^3$), was cut from a 50 g boule and
used in this study. Some general properties of CHC
crystals are listed in Table  \ref{t:chcpro}.
\begin{table}[!ht]
\begin{center}
\caption{Some general properties of CHC crystal scintillators.}
\begin{tabular}{l  |r}
  \hline
  \hline\BottomSpace
    Effective atomic number       &	58\\
    Density (g/cm$^3$)	          &    3.9\\
    Melting point	($^\circ$C)   &       820\\
    Crystal structure	          &      cubic\\
    Wavelength of emission (nm)   &400--430 \\
    Average decay time ($\mu s$)  &       4--5\\
  \hline
  \hline
\end{tabular}
\label{t:chcpro}
\end{center}
\end{table}

\subsection{The detector}\label{p:detector}
The experiment  was carried out at the STELLA  (SubTErranean Low Level Assay)  facility of the LNGS \cite{STELLA}.
The CHC crystal scintillator  was coupled with a 3-inch low-radioactivity photomultiplier (PMT, Hamamatsu R6233MOD), and  placed above the end-cap
\begin{figure}[!htpb]
  \centering
  \includegraphics[width=0.4\textwidth]{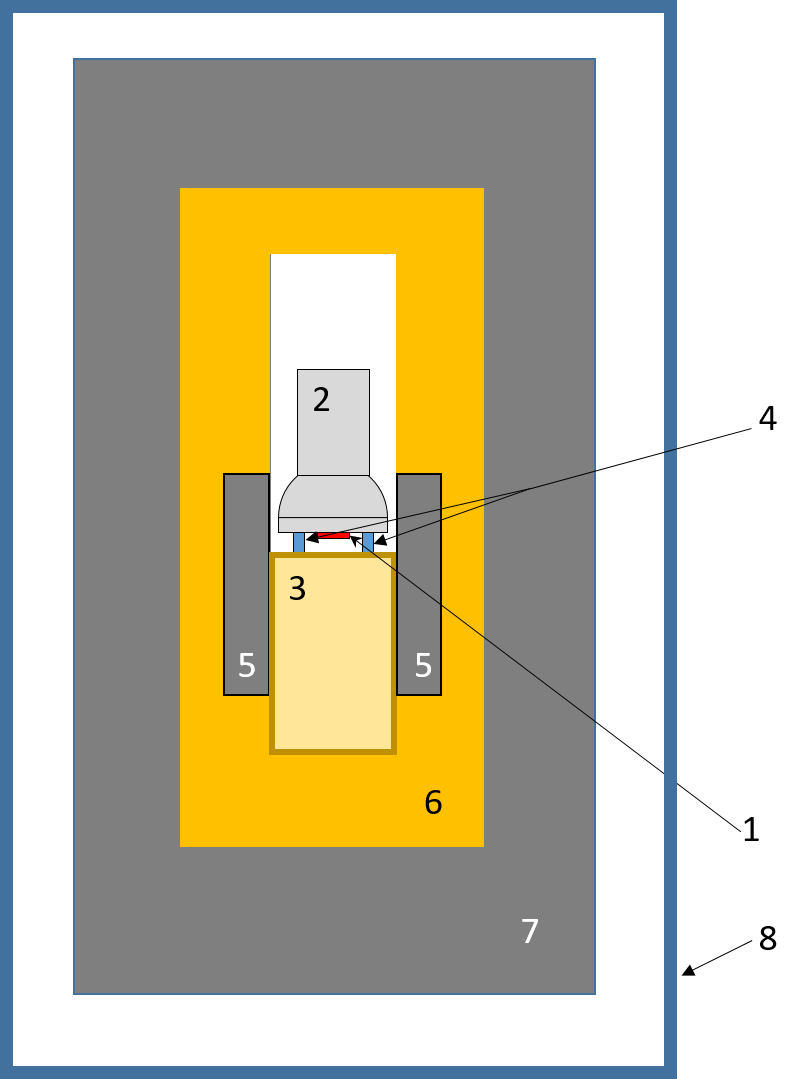}
  \caption{Schematic cross-sectional view of the experimental set-up (not in scale). There are shown the CHC crystal scintillator (1) coupled with a
3 inches PMT (2), the HP-Ge detector (3), which is separated by a cylindrical teflon ring (4). They are completely
surrounded by a passive shield  made by archaeological Roman lead (5), high purity copper (6), low radioactive lead (7).
The whole set-up (with the exception of the cold finger for the HP-Ge detector)  is enclosed in
a plexiglas box (8) continuously flushed with HP-N$_2$ gas.}\label{fig:schema}
\end{figure}
 of the ultra-low
background HP-Ge $\gamma$ spectrometer GeCris (465 cm$^3$).
Due to the longitudinal extent of the CHC detector and of its PMT, the top part of the usual GeCris shield has been slightly re-arranged; no other modifications have been applied to the shield.
A schematic cross-section  of the experimental set-up is shown in Fig. \ref{fig:schema}.

The surface of the CHC crystal was covered with  PTFE tape to improve the light collection.
The passive shield  was assembled (from the external to the internal part) with low radioactivity
lead ($\sim$ 25 cm),  high purity copper ($\sim$ 5 cm), and, in  the inner-most part, with archaeological Roman-age lead ($\sim$ 2.5 cm).
The whole set-up -- with the exception of the cold finger for the HP-Ge detector -- was contained inside a plexiglas box and continuously flushed with high purity (HP) nitrogen gas
to exclude radon close to the detectors.

\begin{figure}[!ht]
  \centering
  \includegraphics[width=0.49\textwidth]{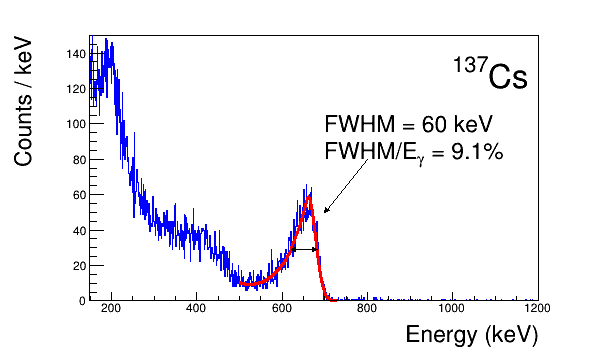}
  \includegraphics[width=0.49\textwidth]{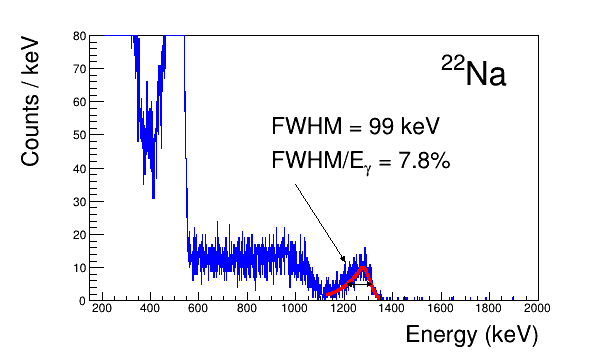}
  \caption[Caption for LOF]{Typical energy distributions collected with the CHC detector for $\gamma$'s from $^{137}$Cs and $^{22}$Na
sources. The fits of the 662 keV and 1275 keV peaks (using the asym--Gaussian function as shape
adopted in Ref. \cite{gausAsim}) are superimposed. The double arrows indicate the FWHM of the two fitted peaks.}\label{fig:csns}
\end{figure}
The signals from the PMT and the HP-Ge were acquired using a CAEN DT5720B digitizer taking 250 MSamples/s;
these signals  were recorded in a time window of 50 $\mu s$. An event-by-event data acquisition system stored the pulse shapes
of the events. The idea was to use, as described in Section \ref{p:t12}, a coincidence logic between the CHC and the HP-Ge detector, to study also the $\alpha$ decay of Hf isotopes to the first excited level, when also a $\gamma$--ray is emitted and could in principle be measured with  the HP-Ge detector. In this way, any possible signal would be directly pointed out from the data.
Thus, hereafter -- unless otherwise stated -- the energy spectra  of events occurred in the CHC detector  and  in  anticoincidence  with the HP-Ge detector are considered.

The energy calibration and resolution of both detectors CHC and HP-Ge  were determined  using
$\gamma$ calibration sources with peak energies  59.5 keV ($^{241}$Am), 511.0 keV ($^{22}$Na), 661.7 keV ($^{137}$Cs) and  1274.5 keV ($^{22}$Na).
In particular, the energy resolution of the CHC detector is:
FWHM(keV) $ = 0.53(5)\times{E}^{0.73(2)}$, where $E$ is in keV.
Fig. \ref{fig:csns} shows the typical energy distributions measured by the CHC detector for $\gamma$'s from $^{137}$Cs and $^{22}$Na sources; there the fits of the 662 keV and of the 1275 keV peaks performed by adopting the asym--Gaussian function as shape (see Ref. \cite{gausAsim}) are superimposed. In fact, in certain cases an asymmetric shape can be expected,
due to either e.g. a non-uniformity of the light collection in the detector volume or to K-escape peaks;
for a complete discussion see Refs. \cite{metAsGgray,studioCsI,semicond,birks}.
\begin{figure}[!ht]
  \centering
  \includegraphics[width=0.9\textwidth]{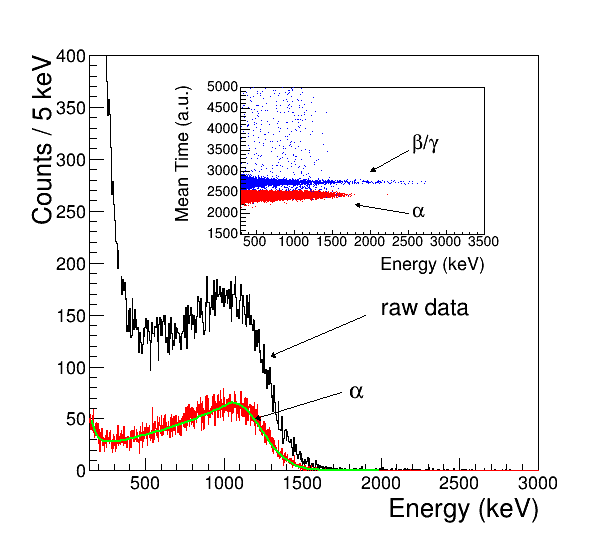}
  \caption{Spectrum of the $^{241}$Am acquired  by the CHC  scintillator. In the inset the PSD plot of the
    $\beta/\gamma$ and $\alpha$ events is reported; see text. In red the $\alpha$ events, in blue the $\beta/\gamma$ events and in black the raw data.
    The superimposed  fit (light green online) is performed using a non-Gaussian function as proposed in Ref. \cite{gausAsim} for the peak and an exponential
    function to describe residual $\beta/\gamma$ events surviving the PSD selection. }\label{fig:am241alfa}
\end{figure}
To monitor the stability of the energy scale, the data were divided in several runs and the stability in time of the energy spectra
was verified.

Moreover, a preliminary dedicated energy calibration of the CHC crystal scintillator was done deep-underground with  a $^{241}$Am $\alpha$ source  in air to obtain a first estimate of the quenching factor (Q.F.)\footnote{The Q.F. describes the response of a scintillator to heavy ionizing particles; in detail, it is the ratio between the detected energy in the energy scale measured with $\gamma$ sources and the actual energy of the heavy ionizing particle.
The Q.F. is an experimental feature of a detector and it is strongly dependent on its nature,  mechanisms of detection and impurities.} of the CHC.
This source was collimated and placed in front of the  CHC detector at a distance of about 1.5 cm. Fig.~\ref{fig:am241alfa}
shows the spectrum of the $^{241}$Am $\alpha$ source acquired  with the CHC  scintillator. In the inset
the PSD of $\beta/\gamma$ and $\alpha$ events  (see section \ref{p:psd}) is shown; the  population with smaller mean time  (red online)
is  mostly due to external $\alpha$'s.
The distribution of the alpha events was fitted with a background model made of an exponential function (to describe residual $\beta/\gamma$ events) plus an asymmetric Gaussian (asym-Gaussian) function  for  the peak\footnote{Besides what mentioned above in the text, the asym-Gaussian shape of the peaks in the case of external alpha's is also due to the partial absorption of the alpha energy in the medium before the CHC.}.
The fit result (light green on-line in Fig.  \ref{fig:am241alfa}) provides an energy peak centroid at $1061(12)$ keV (energy  in $\gamma$ scale) which -- considering
also the $\sim$ 1.5 cm of air between the source and the detector -- gives  a Q.F. $\sim$ 0.4 at that energy.

\section{Low background measurements of the CHC crystal}
The PSD between $\beta/\gamma$ and $\alpha$ particles, and the time-amplitude analysis
of the fast sub-chains of decays from the $^{232}$Th family
 were applied in order to
evaluate the radioactive contamination of the CHC crystal scintillator and the response of
the detector to $\beta/\gamma$ particles.
In particular, the data of the radioactive contamination of the CHC crystal scintillator was used to build a model of the
background in order to derive an estimate of the half-lives of the $\alpha$ decays of the Hf-isotopes,
especially for the $\alpha$ decay of  $^{174}$Hf.

\subsection{Measurements of residual contaminations by ICP--MS and by HP-Ge spectrometry}\label{p:nisi}
In Table~\ref{t:Q} the isotopic composition of $^{\mbox{nat}}$Hf according to  literature is shown. However, the isotopic
\begin{table}[!ht]
\begin{center}
\caption{Isotopic composition of $^{\mbox{nat}}$Hf measured in a sample of the CHC crystal  by ICP-MS.}
\begin{tabular}{c  r}
  \hline
  \hline\BottomSpace
  Isotope     &	   Abundance (\%)       \\
  \hline\TopSpace\BottomSpace
    $^{174}$Hf  &	~ ~ ~0.156(6)\\
    $^{176}$Hf	&     ~ ~5.18(5)\\
    $^{177}$Hf	&       18.5(1)\\
    $^{178}$Hf	&       27.2(1)\\
    $^{179}$Hf	&       13.9(1)\\
    $^{180}$Hf	&       35.2(2)\\
  \hline
  \hline
\end{tabular}
\label{t:iso}
\end{center}
\end{table}
abundance of   $^{174}$Hf is known with  poor precision.
To overcome this problem, a sample of the CHC crystal, 11.8(1) mg, was analysed by means of ICP-MS. The measured abundances are summarized in Tab \ref{t:iso}; in particular, the obtained isotopic abundance of   $^{174}$Hf is 0.156(6)\%,  considerably increasing the precision of the measurement.
This has been obtained by dissolving and diluting  the CHC sample to 100 $\mu$g/l of Hf. In this way, all the Hf isotopes were measured using the ``digital mode" detector avoiding issues due to the Digital/Analogic detector cross calibration (ICP-MS mod 7500a by Agilent technologies was used for isotope composition determination).
In general, when applying high dilution factor before ICP-MS measurements, the isobaric interferences related to the matrix are significantly reduced. Focusing our attention on the mass window of the $^{174}$Hf, it is affected by the $^{174}$Yb signal. But in the CHC sample the presence of the Yb was excluded by monitoring the $^{173}$Yb, which in the measured solution was very low and comparable to the procedural blank one.
The values reported in table 3 are the average of five replicates and their uncertainties are computed with 68\% confidence level.
All  values in the table are in agreement with those reported in literature.

Table  \ref{rad-pur} shows  the level of impurities measured by  ICP-MS. These results have been considered in order to identify the possible presence of  isotopes interfering with   the goal of present work (see later).

\begin{table}[!ht]
\begin{center}
\caption{Concentrations of  trace contaminants in the CHC crystal as measured by ICP-MS analysis.
The limits are at 68\% C.L.}
\begin{tabular}{c  r}
  \hline
  \hline\BottomSpace
  Nuclide   &    	   Concentration (ppb) \\
  \hline\TopSpace\BottomSpace
    $^{144}$Nd  &  $<$2.4             \\
    $^{147}$Sm  &   0.6(1)          \\
    $^{148}$Sm  &   0.4(1)         \\
    $^{151}$Eu  &   19(7)           \\
    $^{152}$Gd  &  $<$0.02            \\
    $^{180}$W   &  $<$0.4             \\
    $^{184}$Os  &  $<$0.003           \\
    $^{186}$Os  &  $<$0.25            \\
    $^{190}$Pt  &  $<$0.02            \\
    $^{209}$Bi	&  $<$2               \\
  \hline
  \hline
\end{tabular}
\label{rad-pur}
\end{center}
\end{table}

The radioactive  contaminations in the CHC crystal were also been measured  using the ultra-low background HP-Ge $\gamma$
spectrometer GeCris of the STELLA facility at LNGS in 841 h of data taking. The results are shown in Table  \ref{rad-cont}.
\begin{table}[!ht]
\begin{center}
\caption{
Radioactive contaminations of the CHC crystal measured with the ultra-low background HP-Ge $\gamma$ spectrometer GeCris of the STELLA facility at LNGS. The uncertainties associated to the measured values are 1$\sigma$, while the upper limits are given at 95\% confidence level (C.L.).
}
\begin{tabular}{l c  r}
\hline
\hline
 Chain & Nuclide & Activity (mBq/kg) \\
 \hline\TopSpace\BottomSpace
           & $^{40}$K              & 0.4(1)$\times 10^{3}$\\
           & $^{44}$Ti             & 10(4)\\
           & $^{60}$Co             & $<$25\\
           & $^{137}$Cs            & 0.74(8)$\times 10^{3}$\\
           & $^{132}$Cs            & $<$15\\
           & $^{134}$Cs            & 79(8)\\
           & $^{181}$Hf            & $<$11\\
           & $^{190}$Pt            & $<$20\\
           & $^{202}$Pb            & $<$9.1\\
\hline\TopSpace\BottomSpace
$^{232}$Th & $^{228}$Ra            & $<$12\\
           & $^{228}$Th            & $<$3.6\\
\hline\TopSpace\BottomSpace
$^{238}$U  & $^{226}$Ra            & $<$23\\
           & $^{234}$Th            & $<$0.80\\
           & $^{234m}$Pa           & $<$0.48\\
\hline\TopSpace\BottomSpace
$^{235}$U  & $^{235}$U             & $<$14\\
 \hline
 \hline
\end{tabular}
\label{rad-cont}
\end{center}
\end{table}

There is no evidence of radionuclides  from the natural decay chains of $^{235}$U, $^{238}$U and $^{232}$Th in the CHC crystal; the corresponding limits have been set at level of a few mBq/kg. The limits on activities of other commonly observed nuclides are also shown in Table  \ref{rad-cont}. However, the CHC crystal is more significantly contaminated with man-made $^{137}$Cs (0.74(8) Bq/kg) and $^{134}$Cs (79(8) mBq/kg). In the case of the cosmogenic radionuclides   $^{132}$Cs and $^{181}$Hf the upper limits are  at the level of 15 mBq/kg and 11 mBq/kg, respectively.

Among the isotopes  listed above, $^{134}$Cs and $^{137}$Cs are rather long-living nuclides with half-lives of 2.0652(4)~y and 30.08~y, respectively; while the other isotopes have relatively short half-life values (T$_{1/2}$ = 6.480(6)~d for $^{132}$Cs, and T$_{1/2}$ = 42.39(6)~d for $^{181}$Hf) making them not problematic for exploring rare decays occurring in Hf isotopes. In a future low-background experiment with CHC detectors, it  necessary to carefully check the initial CsCl compound and find a supplier whose material contains a minimal amount of $^{137}$Cs/$^{134}$Cs. Moreover, to reduce the contamination of the cosmogenic nuclides one should avoid the transportation of produced CHC crystals  by airplane. Finally, it will be of  benefit  storing  CHC crystals  underground  for one-two months before starting  any low-background experiment.
The above mentioned nuclides were present, in traces, also  in a previous work concerning  CHC crystals,  roughly having the same radioactive level  as measured here \cite{nagornyNIMAcontaminanti}. However, the crystal of our work contains significantly higher contamination of $^{40}$K at the level of 0.4(1)~Bq/kg, and $^{44}$Ti with activity of 10(4)~mBq/kg. These radionuclides were not observed in earlier measurements with CHC crystal on the HP-Ge detector and require further studies.
However, it is worth to note that in the paper \cite{sri12}, the GeCris was used to measure SrI$_2$(Eu) and
a contamination of the HP-Ge set-up by $^{44}$Ti (see fig. 9 and text of Ref. \cite{sri12}) was detected. Thus, the same origin could be present in both cases.

\subsection{Pulse-shape discrimination between $\beta/\gamma$ and $\alpha$  particles}\label{p:psd}
\begin{figure}[!ht]
  \centering
  \includegraphics[width=0.9\textwidth]{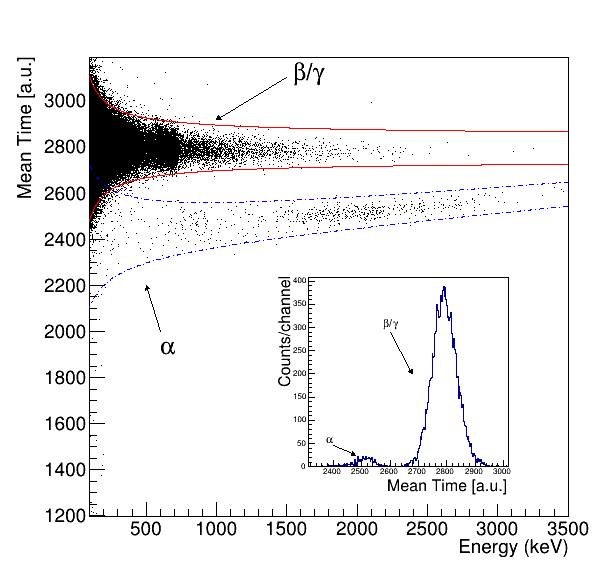}
  \caption{Mean time (see text) versus energy for the low background data accumulated over
2848 h with the CHC detector. The $x$-sigma intervals,  ($x=2.57584$ containing 99\% of events),
around the mean time values corresponding to
$\beta/\gamma$ and $\alpha$  particles are shown (on-line: red solid lines and blue dashed lines, respectively).
(Inset) Distribution of the mean times for the events with
energies in the range of (0.4 -- 3.0) MeV.}\label{fig:psd}
\end{figure}

\begin{figure}[!ht]
  \centering
  \includegraphics[width=0.9\textwidth]{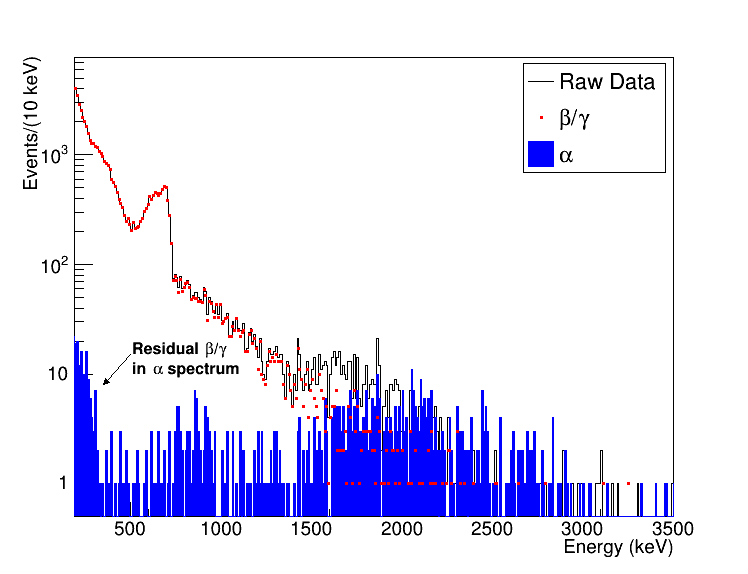}
  \caption{Energy spectrum measured by the CHC detector over 2848 h live-time (raw data; black histogram) and spectra of $\beta/\gamma$ (red dots on-line) and of $\alpha$ events selected by PSD (blu histogram on-line).}
\label{fig:rawAB}
\end{figure}

Scintillation signals from events of different origin ($\alpha$ particles; $\gamma$ quanta or $\beta$ particles) can show different time profiles depending on the detector capabilities; this can be used
to discriminate among them. In our case, we have used a pulse-shape discrimination (PSD)
technique based on the event mean time
(see e.g. \cite{nagornyNIMA,BelliBaF}). In particular, the time profile of each event is exploited to calculate its mean time
according to:
\begin{eqnarray}
\langle t \rangle = \left.\sum f(t_k) t_k\right/\sum f(t_k)
\end{eqnarray}
 where the sum is taken over the time channels, $k$, starting from the origin of the
pulse up to 8 $\mu$s. Moreover, $f(t)$ is the digitized amplitude
(at the time $t$) of a given signal.
The scatter plot of the mean time versus energy for the data of the low background measurements is
shown in Fig. \ref{fig:psd}; it  demonstrates the pulse-shape discrimination ability of
the CHC detector. The
distribution of the mean times for the events with energies -- using the $\gamma$ scale -- in the range (0.4 -- 3.0) MeV is
shown in the inset of Fig. \ref{fig:psd}. The spectra of
$\beta/\gamma$ and $\alpha$ events selected by PSD analysis
are given in Fig. \ref{fig:rawAB}.

\subsection{Time--amplitude analysis of $^{228}$Th sub--chain and the derived Q.F.}\label{TTAQF}
The time--amplitude analysis (described e.g. in Ref. \cite{dan95,dan01,bar00})
was used to select the events of the
following decay sub--chain of the $^{232}$Th family:
$^{224}$Ra (Q$_{\alpha}$ = 5789 keV; T$_{1/2}$ = 3.66 d) $\rightarrow$ $^{220}$Rn (Q$_{\alpha}$ = 6405 keV; T$_{1/2}$ = 55.6 s) $\rightarrow$ $^{216}$Po (Q$_{\alpha}$ = 6906 keV; T$_{1/2}$ = 0.145 s) $\rightarrow$ $^{212}$Pb.

To select these decays, firstly, we search for the pair of $\alpha$’s ($^{220}$Rn -- $^{216}$Po) and after we come back to search for the $\alpha$ from $^{224}$Ra.
 In fact, each $\alpha$ event with energy in the interval (1.4 -- 3.0) MeV
($\gamma$ scale\footnote{The Q.F. of the $\alpha$ events in the CHC has been preliminary estimated in  Section \ref{p:detector} and supported by the data in Ref. \cite{nagornyNIMA}. Taking into account this estimation  the $\alpha$ particles of $^{220}$Rn are expected to fall in this energy region.})
was used as trigger to search – in the same energy interval -- for a second $\alpha$ event ($^{216}$Po)   in the subsequent time interval  (0 -- 1) s (total  efficiency 99.2\%).  This pair of $\alpha$’s ($^{220}$Rn -- $^{216}$Po) was used as trigger to search back for a third $\alpha$ from $^{224}$Ra.
In this latter case, the time interval (1 -- 112) s and the same energy interval as above were considered (total efficiency 74.0\%). Assuming the secular equilibrium of this sub--chain, an average activity
of $^{228}$Th in the CHC crystal scintillator has been estimated: 100(50) $\mu$Bq/kg.
In a data set accumulated over 2176 h with the CHC detector, the energies of the $\alpha$ peaks of  $^{224}$Ra, $^{220}$Rn and $^{216}$Po, selected by the described time-amplitude  analysis, are 2260(200) keV, 2540(200) keV, 2780(240) keV ($\gamma$ scale), respectively.

According to the result of the time--amplitude analysis, the Q.F.
of the used CHC scintillator to $\alpha$ particles at the energies of  $^{224}$Ra, $^{220}$Rn and $^{216}$Po $\alpha$ decays  is 0.39(4),
0.40(3), 0.40(3), respectively.
\begin{figure}[!ht]
  \centering
  \includegraphics[width=0.9\textwidth]{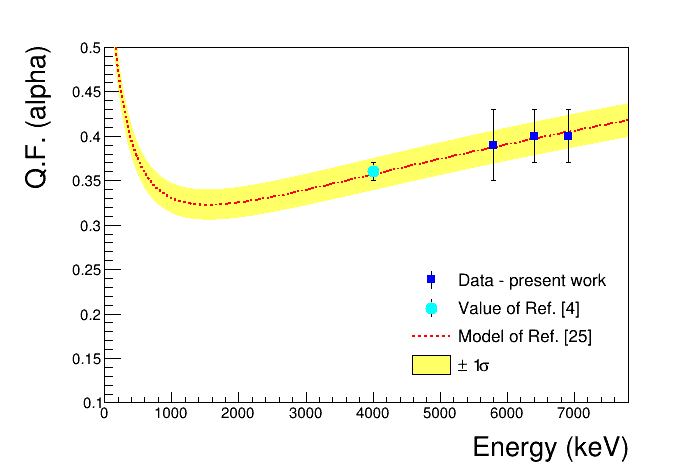}
  \caption{Dependence of the Q.F. on the energy of the $\alpha$ particles measured by the internal $\alpha$ decays of $^{224}$Ra, $^{220}$Rn, $^{216}$Po of the CHC crystal (blue points). The model  obtained as global fits of these
data points following the prescription of Ref. \cite{modelloVIT} are also reported (red dotted line) with its $1\sigma$ fit uncertainty (yellow filled band); the Q.F. for a CHC
scintillator measured in Ref. \cite{nagornyNIMA} is also shown (cyan dot line).}\label{fig:que}
\end{figure}
Fig. \ref{fig:que} shows these Q.F.  values and the global fit (red dotted line) following the prescription of Ref. \cite{modelloVIT},  showing that an $\alpha$ particle
produces  about  1/2.5-th of the light produced by $\gamma$ quanta in the energy range (1--8) MeV. Note that hereafter the energy is in $\alpha$ energy and not in $\gamma$ scale.
In conclusion, considering the model of Ref. \cite{modelloVIT} the Q.F. at 4 MeV  is 0.36, exactly the same Q.F. value measured in Ref. \cite{nagornyNIMA}.
In the following, the fit result according the model of Ref. \cite{modelloVIT} will be used.

\subsection{Identification of Bi--Po events}\label{BIPO}
The search for the fast decays $^{214}$Bi (Q$_\beta$ = 3270 keV, T$_{1/2}$ = 19.9 m) $\rightarrow$ $^{214}$Po (Q$_\alpha$ = 7834 keV,
T$_{1/2}$ = 164 $\mu$s) $\rightarrow$ $^{210}$Pb (in equilibrium with $^{226}$Ra from the $^{238}$U chain) and the fast decays
$^{212}$Bi (Q$_\beta$ = 2252 keV,
T$_{1/2}$ = 60.55 m) $\rightarrow$ $^{212}$Po (Q$_\alpha$ = 8954 keV, T$_{1/2}$ = 0.299 $\mu$s) $\rightarrow$ $^{208}$Pb
was performed with the help of the pulse-shape analysis of the double pulses within the same time window (50 $\mu$s).
We found 11 events; 6 events have a time interval between the starting of the $\beta$ pulse and that of the $\alpha$ pulse
($\Delta T_{BiPo}$) less than 2 $\mu$s. The probability to have a $^{214}$Bi -- $^{214}$Po event in the windows  (0.024 -- 2) $\mu$s\footnote{The lower limit is necessary to be safe when distinguishing two consecutive pulse profiles in the same acquisition window.}
is 0.83\%, while the detection efficiency in the same time window for  $^{212}$Bi -- $^{212}$Po events is 93.6\%. In the case of   $^{214}$Bi -- $^{214}$Po events
the time window  (2 -- 50) $\mu$s has to be considered, and the corresponding efficiency  is 18.2\%.  Taking into account all this
information  one can estimate the activity of $^{226}$Ra in the CHC crystal: $(0.39^{+0.12}_{-0.13})$ mBq/kg, and that of $^{228}$Th:
$(91^{+25}_{-27})$ $\mu$Bq/kg. Thus, the latter is in good agreement with the result of the time-amplitude analysis.

\section{Results on the $\alpha$ decay of naturally occurring Hf isotopes}\label{p:t12}
\begin{figure}[!ht]
  \centering
  \includegraphics[width=\textwidth]{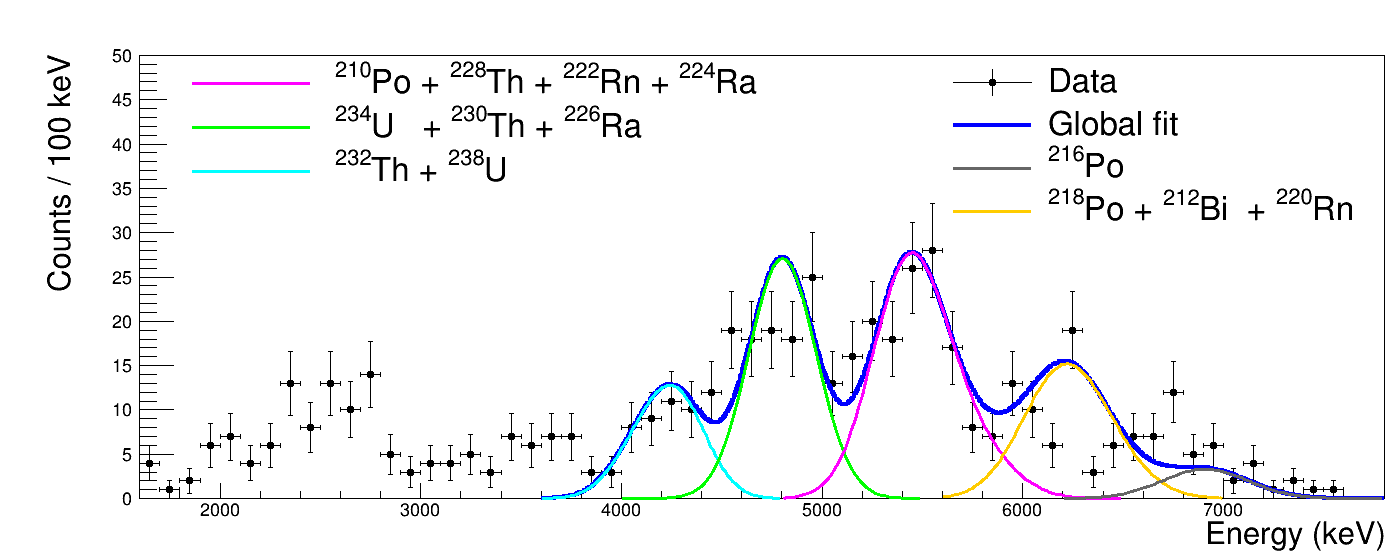}
  \caption{Energy spectrum of the $\alpha$ events selected by PSD
from the data of the low-background measurements with the CHC crystal scintillator over 2848 h. The fit of the data by the model built from $\alpha$ decays of
$^{238}$U and $^{232}$Th with daughters is shown by blue solid line (individual
components of the fit are shown too).  The energy scale is in $\alpha$ energy having considered the Q.F. discussed in Section
\ref{TTAQF}.}\label{fig:spettroalfa2}
\end{figure}

The spectrum of the $\alpha$ events, selected using the PSD analysis, is presented in Fig. \ref{fig:spettroalfa2}; there the energy scale is in $\alpha$ energy having considered the
Q.F. model of Ref. \cite{modelloVIT} as discussed in Section \ref{TTAQF}.

The $\alpha$ energy spectrum below 4 MeV will be discussed later; while the $\alpha$ energy spectrum above 4 MeV has been fitted by using a model which includes
the $\alpha$ peaks\footnote{In this case the peak shapes are considered Gaussian to simplify the fit process and to reduce the number of free parameters (also taking into account  the relative low number of histogram bins with respect to the number of peaks).} of $^{232}$Th, of $^{238}$U and of their daughters in order to study these contaminants.
Besides this, this analysis is useful  to  have an additional check of the adopted Q.F. model. Later, we will focus the analysis in the energy range of interest to study the Hf $\alpha$ decays.
The equilibrium of the
$^{232}$Th and $^{238}$U chains is assumed to be broken in the CHC crystal\footnote{The equilibrium can be broken in the
chains because of the different chemical properties of the nuclides in U/Th chains and of the relatively large half-lives of some nuclides in the chains.}; therefore, the activities
of the following nuclides and related sub-chains (father--last daughter in the following): $^{232}$Th--$^{228}$Ra, $^{228}$Th--$^{208}$Pb and $^{238}$U--$^{234}$U, $^{234}$U--$^{230}$Th, $^{230}$Th--$^{226}$Ra, $^{226}$Ra--$^{210}$Pb, and of the sub-chain
$^{210}$Pb--$^{210}$Bi--$^{210}$Po--$^{206}$Pb,  are  free parameters of the fit.
The energy resolution takes into account the above-mentioned dependence on energy, as measured by calibration.
Moreover, the activities of $^{228}$Th and $^{226}$Ra were fixed to results
of the time-amplitude and of the double pulse (see Sections \ref{TTAQF} and \ref{BIPO}) analyses.

Considering Fig. \ref{fig:spettroalfa2}  one can identify five $\alpha$
structures associated  with the following $\alpha$ peaks \cite{2012Wa38,2017Wa10}:
(i)   $^{216}$Po (Q$_\alpha$ = 6906 keV) ;
(ii)  $^{218}$Po (Q$_\alpha$ = 6115 keV) + $^{212}$Bi (Q$_\alpha$ = 6207 keV) + $^{220}$Rn (Q$_\alpha$ = 6405 keV);
(iii) $^{210}$Po (Q$_\alpha$ = 5407 keV) + $^{228}$Th (Q$_\alpha$ = 5520 keV) + $^{222}$Rn (Q$_\alpha$ = 5590 keV) + $^{224}$Ra (Q$_\alpha$ = 5789 keV);
(iv)  $^{234}$U  (Q$_\alpha$ = 4858 keV) + $^{230}$Th (Q$_\alpha$ = 4770 keV) + $^{226}$Ra (Q$_\alpha$ = 4871 keV);
(v)   $^{232}$Th (Q$_\alpha$ = 4082 keV) + $^{238}$U (Q$_\alpha$ = 4270 keV).
The  fit result,  in the energy interval (3.8--7.8) MeV, is shown in Fig. \ref{fig:spettroalfa2}.
 The fit gives
the activities of $^{238}$U, $^{210}$Po, $^{232}$Th,  $^{226}$Ra and $^{228}$Th  in the crystal, while -- because of a not sufficient  separation between the $^{234}$U and $^{230}$Th
peaks -- one can estimate only the total activity of $^{234}$U and $^{230}$Th; the values are listed in Table  \ref{tab:contaminantiA}.
\begin{table}[!htbp]
\begin{center}
\caption{Measured activities of $^{238}$U, $^{210}$Po, $^{232}$Th,  $^{226}$Ra and $^{228}$Th in the crystal. Because of a not enough good separation between $^{234}$U and
$^{230}$Th
peaks, just the total $\alpha$ activity of $^{234}$U and $^{230}$Th has been derived.}
\begin{tabular}{c c  c }
\hline
\hline
 ~ Chain ~        &  ~Sub-Chain~ & ~Activity (mBq/kg)~ \TopSpace\BottomSpace\\
\hline\TopSpace\BottomSpace
  $^{232}$Th    &  $^{232}$Th & ~0.2(1)  \TopSpace \\
                &  $^{228}$Th & 0.2(1)  \\
\hline\TopSpace\BottomSpace
  $^{238}$U     &  $^{238}$U  & ~0.6(1)   \TopSpace\\
                &  $^{234}$U + $^{230}$Th  & 1.4(2) \\
                &  $^{226}$Ra & 0.2(1)   \\
                &  $^{210}$Po & 1.4(2)   \\
\hline
\hline
\end{tabular}\label{tab:contaminantiA}
\end{center}
\end{table}
Considering all  $\alpha$ events, the total internal $\alpha$ activity in the CHC crystal is at the level of 7.8(3) mBq/kg.

When adopting the claimed half-life of Ref.~\cite{Macfarlane61} (also reported in Table \ref{t:Q}) for the $^{174}$Hf~ $\alpha$ decay, the expected number of events -- within 2848 h of data taking with the used CHC crystal -- is about 1100 counts. Thus, considering that the measured $\alpha$ events are 553(23) in total,  even ascribing all of them to $^{174}$Hf $\alpha$  decay (despite the analysis reported above),  one can safely rule out the result of Ref.~\cite{Macfarlane61}; in fact, even in such an unlike hypothesis, the   T$_{1/2}$ value derived from the present experimental data would be  $4.01(17) \times 10^{15}$ y, i.e.  is about 4.5~$\sigma$ far from the value of Ref. \cite{Macfarlane61}: $T_{1/2} = 2.0(4)\times 10^{15}$. Thus, the T$_{1/2}$ value given in Ref. \cite{Macfarlane61} is safely rejected. Let us now perform a more refined determination of the T$_{1/2}$ value of the $^{174}$Hf~ $\alpha$ decay supported by our data.

Since the Q$_\alpha$ of $^{174}$Hf is 2494.5(2.3) keV, the most problematic dangerous radionuclides -- considering their  natural isotopic
abundances, their Q$_\alpha$ values and their half-lives --  in the search for the $\alpha$ decay of $^{174}$Hf are:
$^{144}$Nd,
$^{147}$Sm,
$^{148}$Sm,
$^{152}$Gd,
$^{186}$Os,
$^{190}$Pt,
$^{209}$Bi. Their main properties are listed in Table  \ref{tab:isotopirompic}.
\begin{table}[!htbp]
\begin{center}
\caption{Properties of the  most problematic radionuclides -- considering their  natural isotopic abundances, their Q$_\alpha$
values, their half-lives, the energy of the emitted $\alpha$ particle -- in the search for the $\alpha$ decay of $^{174}$Hf. In the last column the expected counts, during 2848 hours of data taking with  CHC crystal (6.09(1)~g), calculated according to the mass concentrations reported in Table   \ref{rad-pur} are listed.}
\resizebox{\linewidth}{!}{%
\begin{tabular}{l c  c c c | c}
\hline
\hline\BottomSpace
  Nuclide & Q$_\alpha$                  & T$_{1/2}$         & Isotopic        &  E$_{\alpha}$  & Expected\\
          &    (keV)                    &    (y)            & Abundance       &    (keV)         & Counts\\
          &      \cite{2012Wa38}        & \cite{2012Wa38}   & (\%)\cite{Meija}&                  & \\
 \hline\TopSpace\BottomSpace
$^{144}$Nd &1906.4(17)                  & 2.29(16)$\times 10^{15}$  &  23.798(19)    &  1854.8(17) &$<$0.007\\
$^{147}$Sm &2311.2(10)                  & 1.060(11)$\times 10^{11}$ &  15.00(14)     &  2249.9(10) & 36(6)\\
$^{148}$Sm &1986.9(10)                  & 7(3)$\times 10^{15}$      &  11.25(9)      &  1934.6(10) &$3.6(1)\times10^{-4}$\\
$^{152}$Gd &2204.4(10)\cite{2017Wa10}   & 1.08(8)$\times 10^{14}$   &  0.20(3)       &  2147.8(10) & $<1\times10^{-3}$\\
$^{186}$Os &2820.4(13)                  & 2.0(11)$\times 10^{15}$   &  1.59(64)      &  2761.0(13) &$<6\times10^{-4}$\\
$^{190}$Pt &3252.6(6)                   & 6.5(3)$\times 10^{11}$    &  0.012(2)      &  3185.5(6)  &$<0.1$\\
$^{209}$Bi &3137.3(8)                   & 2.01(8)$\times 10^{19}$   &  100           &  3078.4(8)  &$<4\times10^{-7}$\\
\hline
\hline
\end{tabular}\label{tab:isotopirompic}
}
\end{center}
\end{table}
In particular,  $^{190}$Pt and $^{209}$Bi could contribute in the energy region around 3 MeV, while $^{144}$Nd,
$^{147}$Sm, $^{148}$Sm, $^{152}$Gd and $^{186}$Os
contribute to the energy region of interest.
Taking into account the measured contaminants in Table  \ref{rad-pur}, the expected counts for all isotopes are reported in Table  \ref{tab:isotopirompic}. Thus, only   $^{147}$Sm may give a significant effect with 36(6) counts.

The background model in the energy interval (1.1 -- 3.9) MeV, where the $^{174}$Hf $\alpha$ decay is expected, is made by an exponential function
(to describe residual $\beta/\gamma$ events), and
suitable  asym--Gaussian functions to describe the $\alpha$ decay of $^{147}$Sm (Q$_{\alpha} = 2311.2(10)$ keV), $^{174}$Hf (Q$_{\alpha} = 2494.5(2.3)$ keV) and  the events in the energy range (3.0-3.9) MeV.
These events have been assumed to be degraded $\alpha$  events from possible surface and other contamination.
 The FWHM of the peaks was fixed taking into account the dependence on energy as reported in Section \ref{p:exp}. For the case of  asym--Gaussian used to model the  degraded  $\alpha$ events  the
left tail of the function is used as free parameter, instead  $\sigma$ is limited by the FWHM energy dependence.

\begin{figure}[!ht]
  \centering
  \includegraphics[width=0.45\textwidth]{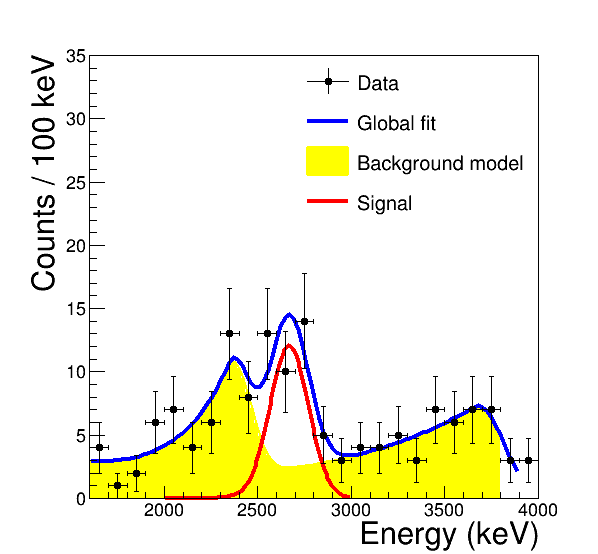}
  \includegraphics[width=0.45\textwidth]{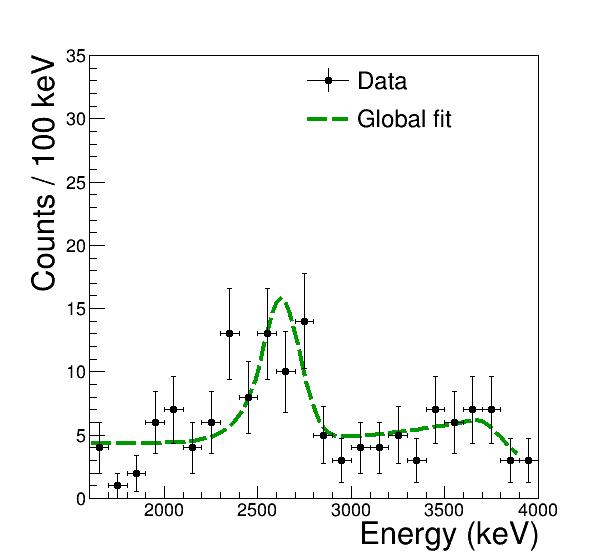}
  \caption{Energy spectrum of $\alpha$ events selected by the pulse-shape discrimination
from the data of low-background measurements with the CHC crystal scintillator over 2848 h. The energy scale is in $\alpha$ energy having considered the Q.F. discussed in Section \ref{TTAQF}.
(\emph{left}) The fit of the data by the model built from $\alpha$ decays of the $^{174}$Hf
(red line) and of the $^{147}$Sm plus degraded alpha particles and an exponential function (to describe residual $\beta/\gamma$ events) is shown (blue solid line online).
The yellow band is the background model with respect to the the signal of the
$\alpha$ decay of the $^{174}$Hf isotope. (\emph{right}) The fit of the data by a modified model similar the previous one but considering just one peak (instead of two) in the energy (2.2 -- 2.6 ) MeV.}\label{fig:spettroalfa}
\end{figure}

 The fit, in the range (1.1 -- 3.9) MeV, provides the area of the peak searched for with
$\chi^2/n.d.f.$=0.87 (P-value = 38.7\%) and $31.7(5.6)$ events. The counts for the peak near 2.3 MeV are 29.5(5.4) in very good agreement with that expected for $^{147}$Sm reported in Table  \ref{tab:isotopirompic}. The Q$_{\alpha}$ of $^{147}$Sm and $^{174}$Hf determined by the fit procedure show a slight variation (for both in the same direction) of the mean value ($\sim 5\%$) consistent with the uncertainty of the adopted Q.F. model discussed in Section \ref{TTAQF}. In addition to the $\chi^2$
test, another independent statistical test has
been applied: the run test (see e.g. Ref. \cite{wald40}); it verifies the hypothesis that the
positive (above the fit value) and negative (under the fit value) data points are
randomly distributed. The lower and upper tail probabilities obtained by the run test
are 94\% and 12\% respectively.

The data in the range (2--3) MeV could be explained in principle by one single peak. In order to study this  possibility we performed a fit of same data as before, but considering only one peak in the energy (2 -- 3) MeV instead of two  (see Fig. \ref{fig:spettroalfa}-\emph{right}); this fit  yields a $\chi^2$ probability of 1.7\%.

To summarize, the analysis support that the data  are  statistically in good agreement with the assumed model, in particular with the assumptions of signals potentially due to   $^{147}$Sm  and $^{174}$Hf  $\alpha$ decays (first and second peak  in Fig. \ref{fig:spettroalfa}-\emph{left} respectively).
To compute the half-life we use the following formula:
\begin{eqnarray}
      T_{1/2}  = \ln2 \cdot N \cdot \epsilon  \cdot t\left/\mbox{S} \mbox{,}\right. \label{f:t12}
\end{eqnarray}
where: i) N is the number of potentially $\alpha$ unstable nuclei; ii) $\epsilon$  is the PSD efficiency that corresponds to 99\% (see above) \footnote{We reasonably assume that all the $\alpha$'s are fully contained in the CHC detector.}; iii) $t$ is the measurement time; iv)
$S$ is the number of events of the effect searched for.
According to all this, the  T$_{1/2}$ value for the $\alpha$ decay of $^{174}$Hf is:
\begin{eqnarray}
    T_{1/2}  = (7.0\pm1.2)\times 10^{16}  \mbox{ y}
\end{eqnarray}

An attempt to improve the experimental sensitivity for  $^{174}$Hf, $^{176}$Hf, $^{177}$Hf, $^{178}$Hf, $^{179}$Hf $\alpha$ decay either to the ground state or to the first excited level has been done. The $^{180}$Hf $\alpha$ decay has not been studied due a too high background in its energy region of interest. The lower limit of half-life is calculated using  formula (\ref{f:t12})
where S is replaced by lim S,
the number of events of the effect searched for which can be excluded at a given
confidence level (in the present work all  half-life limits are given with 90\% C.L.).
\begin{table}[!ht]
\begin{center}
\caption{Half-life lower limits on $\alpha$ decay of  $^{174}$Hf, $^{176}$Hf, $^{177}$Hf, $^{178}$Hf, $^{179}$Hf to the ground state or to the first excited level of relative daughter obtained
from analysis of the data recorded with the CHC crystal scintillator in coincidence with the HP-Ge detector.  All the limits are given at 90\% C.L.}
\resizebox{\linewidth}{!}{%
\begin{tabular}{c | l  c c c c  | c }
\hline
\hline\BottomSpace
 Nuclide Transition                              &  Parent,          &Energy          & Efficiency             & Counts & lim S  & T$_{1/2}$ Limit \\
                                                 &   Daughter        &Level           & $\epsilon$             &        &        & (y)\\
                                                 &    Nuclei         & (keV)          & (\%)                   &        &        &  \\
\hline\TopSpace\BottomSpace
$^{174}$Hf$~\rightarrow~^{170}$Yb                & $0^+\rightarrow2^+$     & 84.3  & 0.11    &0            &2.30    & $1.1\times10^{15}$ \\
\hline\TopSpace\BottomSpace
\multirow{2}*{$^{176}$Hf$~\rightarrow~^{172}$Yb }& $0^+\rightarrow0^+$     & 0     & 99    &$-4.4\pm2.1$ &0.80    & $9.3\times10^{19}$\\
                                                 & $0^+\rightarrow2^+$     & 78.7  & 0.06    &0            &2.30    & $1.8\times10^{16}$\\
\hline\TopSpace\BottomSpace
\multirow{2}*{$^{177}$Hf$~\rightarrow~^{173}$Yb} & $7/2^-\rightarrow5/2^-$ & 0     & 99     &$-4.7\pm2.2$& 0.82   & $3.2\times10^{20}$\\
                                                 & $7/2^-\rightarrow7/2^-$ & 78.6  & 0.06     &0           &2.30    & $7.5\times10^{16}$\\
\hline\TopSpace\BottomSpace
\multirow{2}*{$^{178}$Hf$~\rightarrow~^{174}$Yb} & $0^+\rightarrow0^+$     & 0     & 99     &$3.5\pm1.9$ &6.63    & $5.8\times10^{19}$\\
                                                 & $0^+\rightarrow2^+$     & 76.5  & 0.04     &0           &2.30    & $6.9\times10^{16}$\\
\hline\TopSpace\BottomSpace
\multirow{2}*{$^{179}$Hf$~\rightarrow~^{175}$Yb} & $9/2^+\rightarrow7/2^+$ & 0     & 99     &$-5.4\pm2.3$&0.78    & $2.5\times10^{20}$\\
                                                 & $9/2^+\rightarrow9/2^+$ & 104.5 & 0.65     &0           &2.30    & $5.5\times10^{17}$\\
\hline
\hline
\end{tabular}\label{tab:t12}
}
\end{center}
\end{table}

\begin{table}[!ht]
\begin{center}
\caption{Half-lives on the $\alpha$ decay of Hf isotopes measured in this work in comparison with previous measurements (when possible) and with the theoretical
predictions. All the limits are given at 90\% C.L.}
\resizebox{\linewidth}{!}{%
\begin{tabular}{c | c | c c | ccc }
\hline
\hline\BottomSpace
 Nuclide Transition                               &  Parent,                     &  \multicolumn{5}{c}{T$_{1/2}$}\TopSpace  \\
                                                  &   Daughter                   & \multicolumn{5}{c}{(y)}\BottomSpace             \\
                                                  \cline{3-7}
                                                  &   Nuclei                     & \multicolumn{2}{c|}{Experimental}\TopSpace      & \multicolumn{3}{c}{Theoretical}\TopSpace\\
                                                  &   and its                    &  present work & previous works       &\cite{Abuck91}       &  \cite{poe83}    &  \cite{Teo3}\\
                                                  &   Energy Level (keV)         &  ~             & \cite{danevichchc19} &\multicolumn{3}{|c}{~}\BottomSpace\\
\hline\TopSpace\BottomSpace
\multirow{2}*{$^{174}$Hf$~\rightarrow~^{170}$Yb } & $0^+\rightarrow0^+$, g.s.      & $7.0\pm1.2\times10^{16}$         & $2.0\pm0.4\times10^{15}$\cite{Macfarlane61,T174hf02} &$3.5\cdot10^{16}$  & $7.4\times10^{16}$& $3.5\times10^{16}$\\
                                                  & $0^+\rightarrow2^+$, 84.3      & $\geqslant1.1\times10^{15}$ & $\geqslant3.3\times10^{15}$                   &$1.3\cdot10^{18}$  & $3.0\times10^{18}$& $6.6\times10^{17}$\\
\hline\TopSpace\BottomSpace
\multirow{2}*{$^{176}$Hf$~\rightarrow~^{172}$Yb } & $0^+\rightarrow0^+$, g.s.      & $\geqslant9.3\times10^{19}$& --                                             &$2.5\times10^{20}$& $6.6\times10^{20}$& $2.0\times10^{20}$\\
                                                  & $0^+\rightarrow2^+$,      78.7 & $\geqslant1.8\times10^{16}$& $\geqslant3.0\times10^{17}$                    &$1.3\times10^{22}$& $3.5\times10^{22}$& $4.9\times10^{21}$\\
\hline\TopSpace\BottomSpace
\multirow{2}*{$^{177}$Hf$~\rightarrow~^{173}$Yb}  & $7/2^-\rightarrow5/2^-$, g.s.  & $\geqslant3.2\times10^{20}$& --                                             &$4.5\times10^{20}$& $5.2\times10^{22}$& $4.4\times10^{22}$\\
                                                  & $7/2^-\rightarrow7/2^-$, 78.6  & $\geqslant7.5\times10^{16}$& $\geqslant1.3\times10^{18}$                    &$9.1\times10^{21}$& $1.2\times10^{24}$& $3.6\times10^{23}$\\
\hline\TopSpace\BottomSpace
\multirow{2}*{$^{178}$Hf$~\rightarrow~^{174}$Yb}  & $0^+\rightarrow0^+$, g.s.      & $\geqslant5.8\times10^{19}$& --                                             &$3.4\times10^{23}$& $1.1\times10^{24}$& $2.2\times10^{23}$ \\
                                                  & $0^+\rightarrow2^+$,      76.5 & $\geqslant6.9\times10^{16}$& $\geqslant2.0\times10^{17}$                    &$2.4\times10^{25}$& $8.1\times10^{25}$& $7.1\times10^{24}$\\
\hline\TopSpace\BottomSpace
\multirow{2}*{$^{179}$Hf$~\rightarrow~^{175}$Yb}  & $9/2^+\rightarrow7/2^+$, g.s.  & $\geqslant2.5\times10^{20}$& $\geqslant2.2\times10^{18}$                    &$4.5\times10^{29}$& $4.0\times10^{32}$& $4.7\times10^{31}$\\
                                                  & $9/2^+\rightarrow9/2^+$, 104.5 & $\geqslant5.5\times10^{17}$& $\geqslant2.2\times10^{18}$                    &$2.0\times10^{32}$& $2.5\times10^{35}$& $2.2\times10^{34}$\\
\hline\TopSpace\BottomSpace
\multirow{2}*{$^{180}$Hf$~\rightarrow~^{176}$Yb}  & $9/2^+\rightarrow7/2^+$, g.s.  & --                         & --                                             &$6.4\times10^{45}$& $5.7\times10^{46}$& $9.2\times10^{44}$\\
                                                  & $9/2^+\rightarrow9/2^+$, 82.1  & --                         & $\geqslant1.0\times10^{18}$                    &$4.0\times10^{49}$& $4.1\times10^{50}$& $2.1\times10^{48}$\\
\hline
\hline
\end{tabular}\label{tab:finale}
}
\end{center}
\end{table}

To estimate the  lim S counts of the $\alpha$ decay of Hf isotopes to the ground state, the $\alpha$ spectrum, in the range  (1.1--3.9) MeV, was fitted by a  model composed by the global fit of Fig. \ref{fig:spettroalfa}-\emph{left} plus an asym--Gaussian function for the signal searched for and taking into account the Q$_{\alpha}$ of the transition (see Table  \ref{t:Q}). Table  \ref{tab:t12} reports the respective counts (all compatible with zero) together with the lim S values calculated according the Feldman-Cousins procedure \cite{FelCou}. For these cases, the detection efficiency is reported in Table \ref{tab:t12} and is mostly due to PSD, taking into account that all the  $\alpha$'s are fully contained in the CHC detector. The  lower limits half-life are also  listed. The values span in the range (0.58 -- 3.2) $\times 10^{20}$ y. These are (except for the case of $^{174}$Hf and $^{179}$Hf) ) the first-achieved lower limits of transitions towards g.s. present in literature.\\~

To study the $\alpha$ decay of Hf isotopes to the first excited level, the emitted $\gamma$-rays  detected by the HP-Ge detector in coincidence with the $\alpha$ energy released in the CHC  have been studied. No  signal in coincidence was detected, thus, 2.30 counts at 90\% C.L. have been considered as lim S.
In Table \ref{tab:t12} the detection efficiencies for the coincidences in these decay channels have been estimated by simulating the experiment with EGS4 code \cite{EGS4} and including the PSD efficiency for the $\alpha$ detection; the T$_{1/2}$ limits are spanning in the range $10^{15-17}$~y.
All the results obtained, in comparison with previous measurements and with theoretical
predictions, are summarized in Table  \ref{tab:finale}.
It is worth to note that potentially more stringent limits on such processes, than the caution ones given above, could be reached studying just the alpha spectrum acquired with the CHC detector and exploiting Monte Carlo techniques.

\section {Conclusions}

To study the $\alpha$ decay of  naturally occurring hafnium to the ground state and the first excited state  a CHC crystal scintillator was used in coincidence with a HP-Ge detector in 2848 h of live time. The results rules out the T$_{1/2}$ value of the $\alpha$ decay of $^{174}$Hf  given in Ref. \cite{Macfarlane61}. In particular, we found that the $\alpha$ decay of $^{174}$Hf  to the ground state has been definitely  observed with a  $T_{1/2} = (7.0\pm1.2)\times 10^{16}$~y. This value is in good agreement with the theoretical predictions reported in Table  \ref{tab:finale}.

No signal was detected for $\alpha$ decay of $^{174}$Hf to the first excited state and for $\alpha$ decay of   $^{176}$Hf, $^{177}$Hf, $^{178}$Hf, $^{179}$Hf  either to the ground state or to the first excited level of daughter nuclides.  The  derived  lower limits of the half-life for these decays are reported in Table  \ref{tab:t12}. In particular, the  lower limits for the transitions of $^{176}$Hf$~\rightarrow~^{172}$Yb ( $0^+\rightarrow0^+$) and $^{177}$Hf$~\rightarrow~^{173}$Yb ($7/2^-\rightarrow5/2^-$)  are very close to the theoretical predictions and are (except for the cases of $^{174}$Hf and $^{179}$Hf) the first lower limits of transition between g.s. (see Table  \ref{tab:finale}).

Except for the $\alpha$ decay of $^{174}$Hf for the transition $0^+\rightarrow2^+$, all  other limits  ($\sim10^{16-20}$ y) are very far from the theoretical predictions.

To improve the sensitivity by more than one order of magnitude compared to the present measurements, giving the possibility to observe the alpha decay  $^{176}$Hf$~\rightarrow~^{172}$Yb ($0^+\rightarrow0^+$, g.s.), a larger CHC crystal
$\sim 100$ g (or the enrichment of the CHC crystal with the Hf isotope of interest) could be used and the  $^{147}$Sm contamination reduced by a factor $\simeq$ 10. Moreover, in a future low-background experiment with CHC detectors, it is necessary to reduce as much as possible the amount of $^{137}$Cs/$^{134}$Cs.

We  have also evaluated an average quenching factor for alpha particles, modelled according to the prescription of Ref.  \cite{modelloVIT}, and reported in Fig. \ref{fig:que}; it varies in the range 0.3--0.4. Dedicated measurements, with $\alpha$ sources and  calibrated absorbers, possibly in vacuum, are needed to better study the Q.F.

Finally, the CHC crystal scintillator shows a very interesting PSD capability as shown in Fig. \ref{fig:psd} and  low  U/Th  contamination of few mBq/kg.

\end{document}